\documentclass[useAMS,usenatbib]{mnras} 
\usepackage{graphicx}
\usepackage{amssymb, amsmath}
\usepackage{times}
\usepackage{color}
\usepackage{lscape}
\usepackage[section]{placeins}
\usepackage{breqn}

\usepackage{lineno}

\usepackage{arydshln}

\usepackage{url}

\newcommand{\gtsima}{$\; \buildrel > \over \sim \;$}
\newcommand{\ltsima}{$\; \buildrel < \over \sim \;$}
\newcommand{\simgt}{\lower.7ex\hbox{\gtsima}}
\newcommand{\simlt}{\lower.7ex\hbox{\ltsima}}

\begin{document}


\title{Relativistic distortions in the large-scale clustering of SDSS-III BOSS CMASS galaxies }

\author[Alam et al.] {
    Shadab Alam$^{1,2,3}$ \thanks{Email: salam@roe.ac.uk}, Hongyu Zhu$^{1,2}$, Rupert A. C. Croft$^{1,2}$, Shirley Ho$^{1,2,4,5}$,
    \newauthor
     Elena Giusarma$^{1,2,4,5}$ and Donald P. Schneider$^{6,7}$ \\
    $^{1}$Department of Physics, Carnegie Mellon University, 5000 Forbes Ave., Pittsburgh, PA 15213 \\
    $^{2}$McWilliams Center for Cosmology, Carnegie Mellon University, 5000 Forbes Ave., Pittsburgh, PA 15213 \\
    $^{3}$Institute for Astronomy, University of Edinburgh, Royal Observatory, Blackford Hill, Edinburgh, EH9 3HJ, UK\\
    $^{4}$Lawrence Berkeley National Laboratory (LBNL), Physics Division, Berkeley, CA 94720-8153, USA \\
    $^{5}$Berkeley Center for Cosmological Physics, University of California, Berkeley, CA 94720, USA \\
    $^{6}$Department of Astronomy and Astrophysics, The Pennsylvania State University,University Park, PA 16802 \\
    $^{7}$Institute for Gravitation and the Cosmos, The Pennsylvania State University, University Park, PA 16802
}
    
\date{\today}
\pagerange{\pageref{firstpage}--\pageref{lastpage}}   \pubyear{2015}
\maketitle
\label{firstpage}

\begin{abstract}
General relativistic effects have  long been predicted
to subtly influence the observed large-scale  structure of the universe.
The current generation of galaxy redshift surveys have reached a size 
where detection of such effects is becoming feasible. In 
this paper, we report the first detection of the redshift asymmetry from the cross-correlation function of two galaxy populations which is consistent with relativistic effects.
The dataset is taken from the Sloan Digital Sky Survey DR12 CMASS galaxy sample,
and we detect the asymmetry at the $2.7\sigma$ level by applying a
 shell-averaged estimator to the cross-correlation function. 
Our measurement dominates at scales around $10$ h$^{-1}$Mpc, larger
than those over which the gravitational redshift profile has been recently
measured in galaxy clusters, but smaller than scales for which
linear perturbation theory is likely to be accurate.  The 
detection significance varies by 0.5$\sigma$ with the details of 
our measurement and tests for systematic effects. We have also 
devised two null 
tests to check  for various survey systematics and show that both results are consistent 
with the null hypothesis. We measure the dipole moment 
of the cross-correlation function, and from this the asymmetry is 
also detected, at  the 
$2.8 \sigma$ level. The amplitude and scale-dependence of the clustering asymmetries  are approximately consistent with the expectations of General Relativity and a biased galaxy population, within large  uncertainties. We explore theoretical predictions using numerical  simulations in a companion paper.

\end{abstract}

\begin{keywords}
    gravitation; modified gravity;
    galaxies: statistics;
    cosmological parameters;
    large-scale structure of Universe
\end{keywords}

\section{Introduction}
\label{sec:intro}
The General Theory of Relativity  \citep[GR]{Einstein1916} has been  successfully applied to the prediction of the structure of our Universe. As a theory it  provides a complete account of the gravitational matter-matter and light-matter interactions. Einstein proposed three tests of general relativity, the perihelion precession of Mercury's orbit \citep{Clemence1947}, the deflection of light by the Sun \citep{Dyson1919,Kennefick2007} and the gravitational redshift of light \citep{Pound1959}.GR has been tested against many other observations over last century including post-Newtonian tests of gravity \citep{Nordtvedt1985}, the light travel time delay around massive objects, also known as Shapiro delay \citep{Shapiro1964}, constraints on the strong equivalence principle \citep{Nordtvedt1968a, Nordtvedt1968b}, weak and strong gravitational lensing \citep{2005astro.ph..9252S}, cosmological tests using the growth rate from large scale structure of galaxies \citep{Kaiser87,Alam2016} and the $E_G$ parameter \citep{Zhang2007, Reyes2010, Pullen2015, Pullen2015data}, indirect detection of gravitational waves through pulsar timing \citep{Weisberg1981} and recent direct detection through a binary black hole merger \citep{Abbott2016}. Recently several authors have studied relativistic effects on the large scale structure observed in galaxy redshift surveys \citep{McDonald2009, Yoo2009, Jeong2012, Yoo2012, Croft2013, Yoo2014a, Bonvin2014a,Bonvin2014b}. These papers will hopefully mark the beginning
of a new era testing general relativity by analyzing galaxy clustering with unprecedented precision (for a review see \citet{Yoo2014}). 

The Universe is assumed to be isotropic and hence any statistical property (for example, distribution of galaxies) is expected to be isotropic. Galaxy redshift surveys have made measurements of millions of galaxies in regions of the Universe and analyzed their large-scale clustering properties [2dF: \citet{Colless2003}, 6dF: \citet{Jones2009}, SDSS-III: \citet{Eisenstein2011},  WiggleZ: \citet{WiggleZ}, DEEP2: \citet{Deep2013}, VIPERS: \citet{Garilli2014},  GAMA: \citet{gama2015}]. The two point correlation functions (2PCF) of observed galaxies  in these surveys are far from isotropic due to observational effects. The line-of-sight galaxy distances from the Earth are inferred from the redshift ($z$) of spectral features, assuming a cosmological model. As well
 as the distance, each 
redshift also  contains information on the dynamics (peculiar velocity) and the environment (gravitational potential) of these galaxies. The redshift has three components: the Hubble recession velocity, the peculiar velocity of a galaxy and the local gravitational potential. The observed redshift ($z_{\rm obs}$) is given by
\begin{equation}
z_{\rm obs} = H(z)r/c + v_{\rm pec}/c + z_g,
\end{equation}
where $H(z)$ is the Hubble parameter, $r$ is the true line-of-sight distance, $v_{\rm pec}$ is the peculiar velocity of galaxy, $c$ is the speed of light, $z_g$ is the gravitational redshift. The expression is valid for 
distances $r$ where a linear Hubble relation is a good approximation. The peculiar velocity component of the observed redshift modifies the galaxy two-point correlation function, causing redshift space distortions (RSD). The effect of RSD is manifested in changes in the angle averaged `even ordered multipoles' of the two point correlation function of galaxies, most prominently in the second order multipole (quadrupole) \citep{Hamilton92}. As the peculiar velocities of galaxies are isotropically oriented on average, only the even ordered multipoles of the 2PCF remain non-zero while the odd ordered multipoles vanish. \citet{Peebles1980} presented one of the first discussions of RSD affecting the large scale structure of the Universe. The first linear theory formalism to model RSD was developed by \citet{Kaiser87}. Over the last few decades the analysis of various galaxy redshift   surveys have improved our understanding of large-scale structure. Recent studies of RSD in this context include   \cite{2dFGRS,Blake2011,6dFGRS,Vipers,SDSSLRG2012, Reid12, Ariel13,Beutler13,Reid14,Cullan14,Alam2015, Simpson2016}. 

\begin{figure}
\begin{center}
\includegraphics[width=0.5\textwidth]{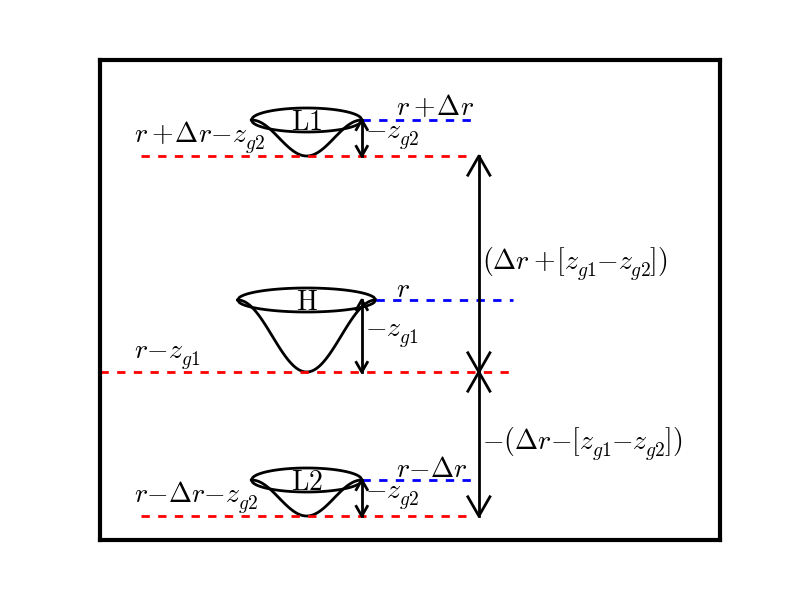}
\caption{An illustration of the symmetry breaking along line-of-sight due to gravitational potential. In this symmetric system of three galaxies the central galaxy (${\rm H}$) is more massive than the other two galaxies (${\rm L1,L2}$). The real space distance of each of the smaller galaxies from the central galaxy is identically $\Delta r$. Each of these galaxies is hosted by a dark matter halo which adds extra gravitational components ($z_{g1}$ and $z_{g2}$) to the measured redshifts.}
\label{fig:illusration}
\end{center}
\end{figure}

The gravitational redshift component of the galaxy redshift is generated
by the environment of the galaxy. Galaxies are known to occupy dark matter haloes which possess strong gravitational potentials ($\phi$). 
The light emitted from a galaxy will experience a gravitational shift $z_g=-\phi /c^2$ as it climbs out of the potential well. This effect is one of the most fundamental predictions of GR. The manifestation of this phenomenon was first observed in a nuclear resonance experiment and reported by \citet{Pound1959}. The gravitational redshift has also been measured in astrophysical systems, e.g; a white dwarf sirius B \citep{Greenstein1971}, in the solar system \citep{Lopresto1991,Takeda2012} and in galaxy clusters \citep{Wojtak2011,Sadeh2015,Jimeno2015}. A formalism for measuring the gravitational redshift in individual galaxy clusters was presented by \citet{Cappi1995}, later it was suggested that stacking large numbers of galaxies could be used to measure the gravitational redshift profile of clusters as a function of scale \citep{Kim2004}.

In this paper, we present measurements of the
 relativistic distortions of galaxy clustering on scales larger than
clusters. We  use the cross-correlation of two galaxy populations with different masses. \citet{McDonald2009} provided the first linear theory formalism to predict the effect of gravitational redshift in the cross power spectrum of two different populations of galaxies. \citet{Croft2013} carried out $N$-body simulations of the effect of gravitational redshift on two population of haloes with different halo masses, revealing that allowing for measurements
on non-linear scales, current galaxy redshift surveys should be able to detect such an effect, and future surveys should be able to provide precise measurements.  \citet{Croft2013} also proposed an estimator to measure the line-of-sight
asymmetry using the cross-correlation function of the two differently biased samples. 
Examining this effect at galaxy cluster scales, \citet{Kaiser2013} and \citet{zhao2013} demonstrated that gravitational redshift is the dominant of several
relativistic distortions of non-linear clustering which should be
considered together. \citet{Bonvin2014b} showed how gravitational
redshift distortion is related to the full general relativistic asymmetry
of the cross-correlation of two populations of galaxies. A measurement
of the cross-correlation dipole (from datasets which included the
SDSS CMASS sample) was made by \cite{Gaztanaga2015}. This
was on large ($r > 20$ h$^{-1}$Mpc) scales where the relativistic distortions
were not measurable but a purely geometrical distortion was seen.

Figure \ref{fig:illusration} illustrates how gravitational redshifts  
act to cause an asymmetry in clustering  (alongside
other relativistic effects). 
Consider a symmetric system of three galaxies where the central galaxy (${\rm H}$) is more massive than the other two galaxies (${\rm L1,L2}$). The distance of each of the smaller galaxies from the central galaxy is identically $\Delta r$. Imagine that each of these galaxies is hosted by a dark matter halo which adds extra gravitational components to the measured redshifts $z_{g1}$ and $z_{g2}$. If we include these effects and then examine the redshift
difference of each lower mass galaxy from the central galaxy, we will find that ${\rm L1}$ is located at a distance $\Delta r +(z_{g1}-z_{g2})$ and ${\rm L2}$ at $\Delta r -(z_{g1}-z_{g2})$  from the central galaxy. The line of sight redshift-space distance of the two galaxies will be equal in the limit $z_{g1}=z_{g2}$. In the scenario when the galaxies reside in haloes of different masses  (different gravitational potentials) the symmetry along the line of sight will break. This effect will produce odd ordered moments in the cross-correlations of galaxies and could be used to measure the different environments of the galaxies. An  important factor to note that in the autocorrelation function is that the pair counts
are symmetric by construction, so no distortions of this type can be 
measured.

The line of sight asymmetry illustrated in Figure \ref{fig:illusration} can also be introduced by other observational effects \cite{Kaiser2013,Bonvin2014b}.
We have studied some of the relativistic effects on target selection (observational systematics) in a companion paper \citet[][in press.]{Alam2016TS} . A study of all such relevant effects using $N$-body simulations is presented in another companion paper \citet[][in press.]{Zhu2016Nbody}.
 We have also investigated the effects of baryons on the gravitational potential and
velocities of galaxies and their impact of relativistic distortions using hydrodynamical simulations in a third companion paper \citet[][in prep.]{Zhu2016Hydro}.

A  outline of this paper is as follows. 
We provide a brief overview of our theoretical model in Section \ref{sec:theory}. Section \ref{sec:data} presents our dataset and Section \ref{sec:measurement} describes the steps and methods used in our measurements.  The results are presented in Section \ref{sec:result}, which include a $3.5\sigma$ detection of the line of sight asymmetry using large scale structure. We also discuss several null tests and systematics tests perform on the data. We conclude in section \ref{sec:discussion} with a summary and a discussion of our measurement.

\section{Theory}
\label{sec:theory}

The existence of gravitational redshift is one of the fundamental 
predictions of GR. As mentioned in Section \ref{sec:intro}, it has been previously studied theoretically and observed experimentally on various scales. 
We use $N$-body simulations to predict the measured signal from gravitational
redshifts and other effects which distort the cross-correlation
function.  The perturbation theory approach  \citep[e.g.][]{McDonald2009, Yoo2009,Bonvin2014a} is valid on large-scales, but  
 non-linear clustering (including the structure of the potential well on galactic and halo scales) is dominant on the scales which are accessible to current observations (\cite[][in press.]{Giusarma2016PT}, in prep.). We therefore use N-body simulations to make predictions for the gravitational potential and velocities of galaxies. The suite of $N$-body simulations and details of how they were used to predict different components of the signal are presented in a companion paper \citet[][in press.]{Zhu2016Nbody}. Here, we briefly describe the simulations and effects included in our theoretical model.

We use $N$-body simulations produced by running the PGadget3 code \citep{springel2001,springel2005}. We use eight realizations of a flat $\Lambda$CDM model with $\Omega_m=0.3$ and $h=0.7$. The simulations are dissipationless, in a periodic box of side length $1000 {\rm h^{-1}Mpc}$ and contain $1024^3$ particles. We use SubFind \citep{springel2001} to generate a catalog of subhalos which can be associated with galaxies. We then observed the subhalo catalog in the parallel line-of-sight approximation including various observational effects.
We also include ``wide-angle'' effects which have been shown to become important on large, linear scales \citep{Gaztanaga2015} based on linear theory prescriptions. The observed line-of-sight position is given as follows:

\begin{equation}
Z_{\rm obs}=Z_{\rm real}+Z_{\rm pec}+Z_g+Z_{\rm TD}+Z_{\rm LC} +Z_{\rm galaxy},
\end{equation}
where $Z_{\rm obs}$ is the final observed distance from the observer in comoving units ($h^{-1}{\rm Mpc}$)  including all the effects we study. From now on, $v$ will denote velocity while $\beta$ denotes the ratio $v/c$, where $c$ is the speed of light. Also, $H=100$ (km/s) / (h$^{-1}$Mpc). The cartesian components of velocity and $\beta$ are indicated using subscripts (e.g., $v_{x,y,z}, \beta_{x,y,z}$).  $Z_{\rm pec}=v_{z}/H$ gives the effect of peculiar velocity 
on the line-of-sight distance, and is the term which causes the usual
redshift distortions (e.g,  \citep{Kaiser87}). The quantity $Z_g=-\phi/(cH)$ is the positional shift caused by the gravitational redshift from the subhalo potential \citep{Cappi1995}. Two more terms introduce additional redshifts that depends on the peculiar velocity: $Z_{\rm TD}=\beta^2c/(2H)$ accounts for the Transverse Doppler effect \citep{zhao2013} and $Z_{\rm LC}=\beta_z^2c/(H)$ the light cone effect \citep{Kaiser2013}. Because the potential well of the stellar component of the galaxy adds to the gravitational redshift, we add the component due to internal structure of galaxy similar to \citet{Cappi1995}, $Z_{\rm galaxy}=10^{-5} \sigma_v^2  \ln(R_{\rm{e}}^{\rm dm}/R_{\rm{e}}^{\rm star})$, where $R_{\rm e}^{\rm dm}$ and $R_{\rm e}^{\rm star}$ are half-mass radius of dark matter and star particles respectively estimated from MBII hydrodynamical simulation \citep{khandai2015,Zhu2016Hydro}.

 After adding all these components to the observed position of each galaxy in the simulation, we construct two populations of subhalos divided by median subhalo mass. The halo masses used to obtain two subsamples accounts for the luminosity distance perturbation as described in section 2.5 and specifically equation 8 of \cite[][in press.]{Zhu2016Nbody}. We measure their auto-correlation functions and define two linear
bias values, $b_{H}$, and $b_{L}$, for the high and low mass halves of the sample respectively. We compute the bias in the usual fashion (see section \ref{sec:mbias} for details)
from the relative scaling of the large scale auto-correlation
functions of subhalos and dark matter in the simulations 
(see \cite[][in press.]{Zhu2016Nbody} for details).

\begin{figure}
\includegraphics[width=0.5\textwidth]{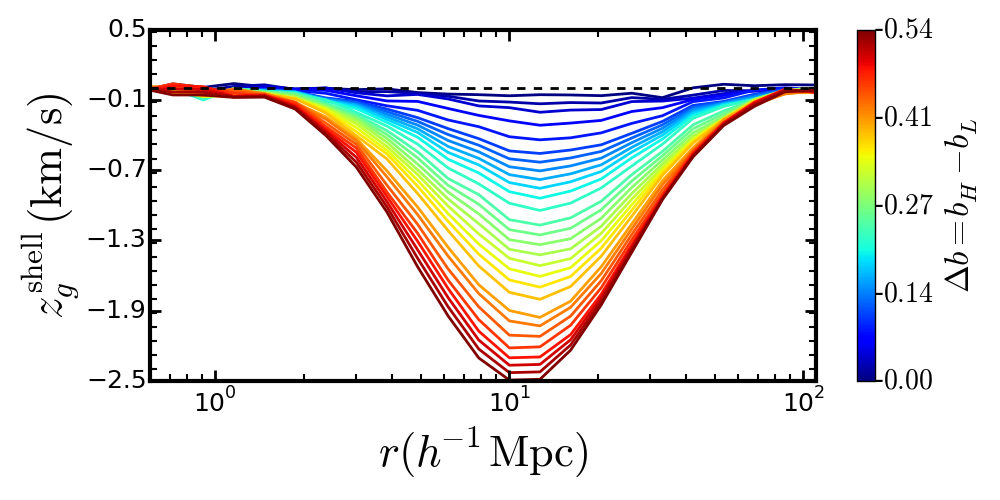}
\includegraphics[width=0.44\textwidth]{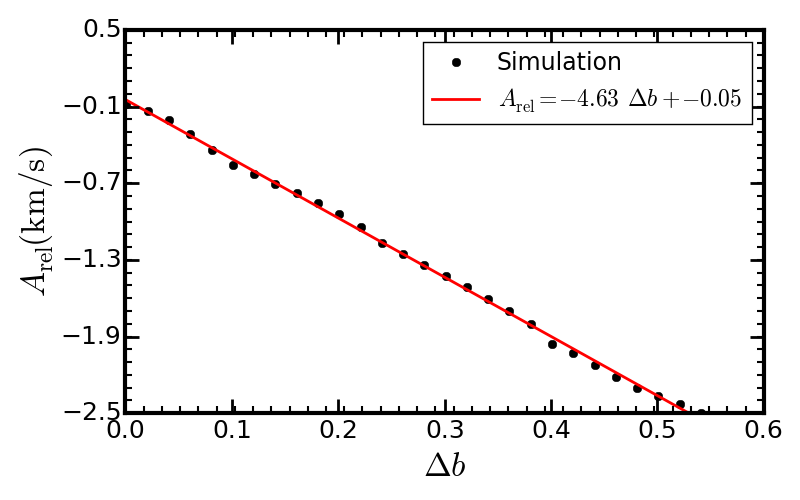}
\caption{
The asymmetry signal in the cross-correlation function of
two galaxy populations measured from $N$-body simulations using the shell
estimator of Equation \ref{eq:shell}. The top panel  shows results for various values of  $\Delta b$, the bias difference between the two 
population (as indicated by the colour bar). The bottom panel presents the amplitude of the shell estimator as the function of $\Delta b$, measured from simulations (black points) and best fit linear model (red line, Equation \ref{eq:gz-model}).}
\label{fig:zg-theory}
\end{figure}

To measure the asymmetry in the cross-correlation function of the two samples we use the shell estimator, proposed by \citet{Croft2013}. This technique is simply 
an estimate of the centroid shift of the correlation function averaged in 
spherical shells, and is similar to the usual gravitational
redshift profile of clusters \citep[e.g.][]{Wojtak2011}.  The shell estimator is measured from the two subsamples as follows:

\begin{equation}
z_g^{\rm shell} (r) = H \sum_{h_1,h_2}^{|r_{12}-r|<\Delta r} [Z_{\rm obs}^{h1}-Z_{\rm obs}^{h_2}]\frac{P_{\beta}(h_1) P_{\beta}(h_2)}{\sum_{h_1,h_2}^{|r_{12}-r|<\Delta r}P_{\beta}(h_1) P_{\beta}(h_2)}
\label{eq:shell}
\end{equation}
where the sum is over all pairs of subhalos ($h_1,h_2$) between the two populations such that the distance between the subhalos $r_{12}$ lies in the radial bin between $r-\Delta r$ and $r+\Delta r$. 
Here $P_\beta=1-1.0*\beta_z$ accounts for the relativistic beaming effect on
galaxy inclusion in the sample by weighting the pair in a fashion which depends on their line of sight velocities \citep{Kaiser2013, Alam2016TS, Irsic2016}. The beaming model is derived in our companion paper \citep[][in press.]{Zhu2016Nbody}  based on observational probability measured for CMASS sample in another companion paper \citep[][in press.]{Alam2016TS}. The above definition of shell estimator is equivalent to the definition given in equation \ref{eq:zg-rtheta}. Lastly we add contribution of wide-angle effect to our model as described in section 2.6 of \cite[][in press.]{Zhu2016Nbody}.

Figure \ref{fig:zg-theory} (top panel) shows measurements of cross-correlation
function asymmetry produced by applying the shell estimator to our $N$-body simulation outputs.
 Lines of different colours show measurements for various values of the large-scale bias difference between the two populations. We have used these simulation results  to develop a parametric phenomenological model for the
shell estimator as follows: 

\begin{equation}
z_g^{model}(r) = \frac{A_{\rm rel}(\Delta b)}{A_{\rm rel}(\Delta b=0.3)} z_g^{theo}( r),
\label{eq:gz-model}
\end{equation}
where $z_g^{theo}$ is measured from simulations for $\Delta b=0.3$. Here $A_{\rm rel}$ is a function of bias difference ($\Delta b=b_{H}-b_{L}$) which determines the amplitude of asymmetry. We choose $\Delta b=0.3$ for normalization because it lies in the middle of the $\Delta b$ range explored in the simulations. We have examined the amplitude as the function of bias difference from the simulations. The scaling of amplitude with bias difference is well described with a linear function. 

\begin{equation}
A_{\rm rel}(\Delta b)= -4.63 (\Delta b ) - 0.05
\end{equation}

The amplitude scaling of this model is motivated by the perturbation theory results; see equations 12 and 30 from \citet{Gaztanaga2015}, which show that the shell estimator at linear scale should be proportional to the bias difference of the two sub-samples. Since we are working at the non-linear scale we naively expect the amplitude of the signal to depend on higher order function of linear bias difference, but the simulations show a linear scaling of amplitude with bias difference. The position of the largest asymmetry depends on the mean bias of the two populations, redshift and the relative contribution of different relativistic effects \citep{CaiGz2016, Alam2016TS}.

Figure \ref{fig:zg-theory} (bottom panel) displays the amplitude of the shell estimator as the function of $\Delta b$, measured from simulations with black points and best fit linear model with a red line (Equation \ref{eq:gz-model}).  Our model captures the variations in measured signal from $N$-body simulations adequately for our purposes.

\section{Data}
\label{sec:data}
To perform our measurement of relativistic distortions in the cross-correlation
function we use observations from Data Release 12 (DR12;\cite{Reid2016,Alam2014}) of the Sloan Digital Sky Survey \citep[SDSS;][]{York2000}. SDSS I, II \citep{Abazajian2009} and III \citep{Eisenstein2011} used a drift-scanning mosaic CCD camera \citep{Gunn1998} to image 14555 square degrees of the sky in five photometric bands \citep{Fukugita1996,Smith2002,Doi2010} to a limiting magnitude of $r <22.5$ using the  2.5-m Sloan Telescope \citep{Gunn2006} at the Apache Point Observatory in New Mexico. The imaging data were processed through a series of software pipelines \citep{Lupton1999,Pier2003,Padmanabhan2008}. \cite{Aihara2011} reprocessed all of the SDSS imaging data as part of Data Release 8 (DR8). The Baryon Oscillation Spectroscopic Survey \citep[BOSS;][]{Dawson2013} was designed to obtain spectra and redshifts for 1.35 million galaxies covering 10,000 square degrees of sky. These galaxies were selected from the SDSS imaging. \citep{Blanton2003b} developed a tiling algorithm for BOSS that is adaptive to the density of targets on the sky. BOSS used double-armed spectrographs \citet{Smee2013} to obtain the spectra and completed observations in spring 2014. BOSS obtained a homogeneous data set with a redshift completeness of more than 97\% over the full survey footprint. The redshift extraction algorithm used in BOSS is described in \citet{Bolton2012}. \citet{Eisenstein2011} provides a summary and \citet{Dawson2013}  provides a detailed description of the survey design.

We use the CMASS sample of galaxies  \citep{Bolton2012} from DR12 \citep{Alam2014}. The CMASS sample has 7,65,433 Luminous Red Galaxies (LRGs) covering 9376 square degrees in the redshift range $0.44<z<0.70$, which correspond to an effective volume of 10.8 Gpc$^{3}$.

\section{Analysis Methodology}
\label{sec:measurement}

\begin{figure}
\begin{center}
\includegraphics[width=0.5\textwidth]{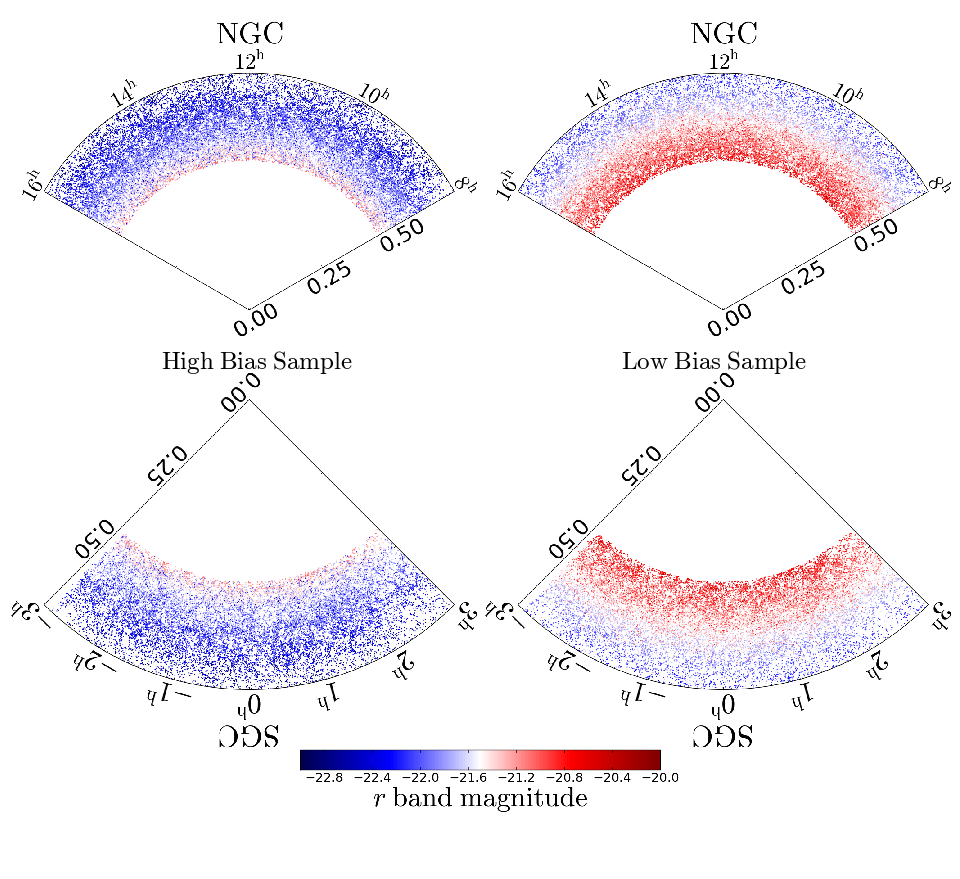}
\caption{The distribution of the galaxies in our CMASS sample. The radial distance
in each segment corresponds to the redshift of a galaxy and the angle corresponds to its right ascension (RA). The colour denotes the $r$ band magnitude of the galaxy. The top segments in each side show the north galactic cap (${\rm NGC}$)  and the bottom  the south galactic cap (${\rm SGC}$) for the two sub-samples. The left panel displays the high bias (more massive) subsample of
the CMASS data and the right panel the lower bias (less massive) samples.}
\label{fig:subsample}
\end{center}
\end{figure}

\begin{figure}
\begin{center}
\includegraphics[width=0.5\textwidth]{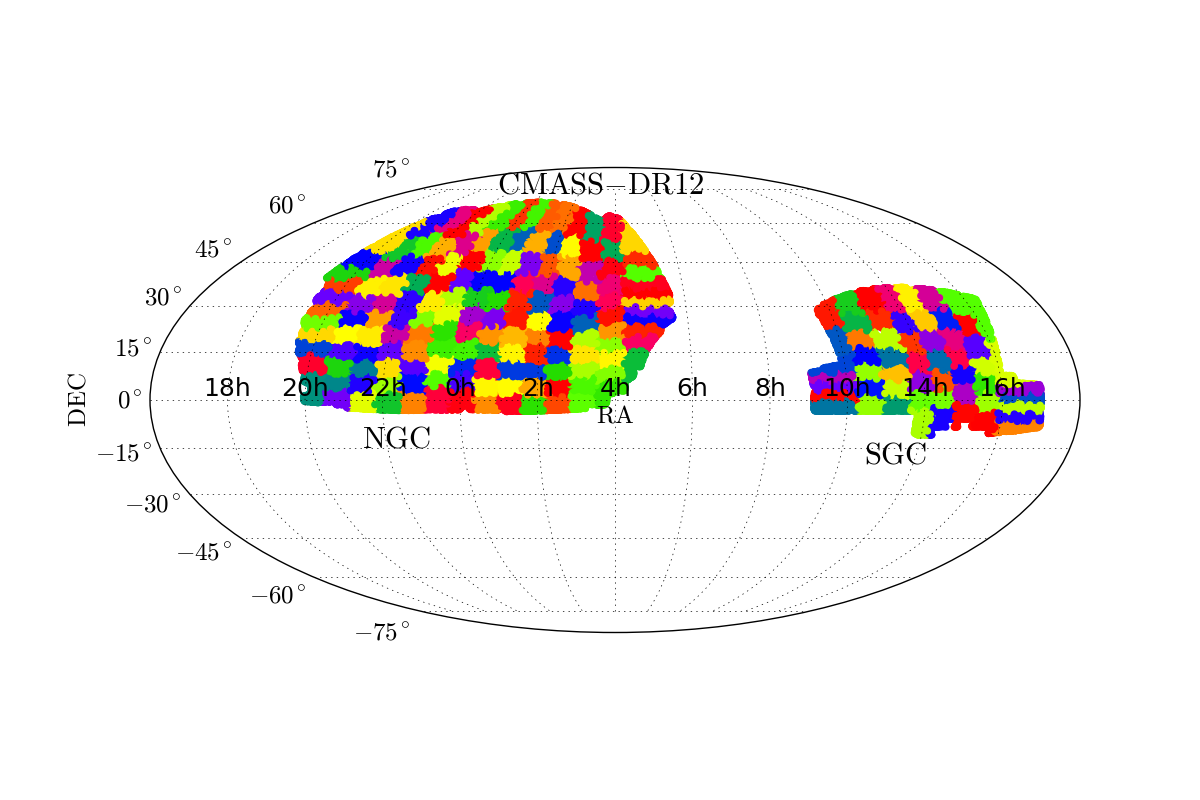}
\caption{The distribution of galaxies in our CMASS sample on the sky. The two seperate regions are the NGC (left) and SGC (right). Different colours correspond to different jackknife regions used in calculation of the covariance matrix. The origin of RA has been shifted by 4 hours towards the left in order to have the SGC appear as a contiguous region.}
\label{fig:jacknife}
\end{center}
\end{figure}

The CMASS sample is used to define two galaxy sub-samples with different biases but same redshift distributions. We measure the auto-correlation
function for each sub-sample, and estimate their linear bias values. We then measure the cross-correlation function of  the two sub-samples. The cross-correlation function is used to detect and quantify the line of sight asymmetry due to relativistic effects using shell estimator of equation \ref{eq:zg-rtheta}. In this section we describe our analysis methodologies in detail.

\subsection{Creating Galaxy sub-samples}
The SDSS CMASS DR12 sample of  galaxies are used create two 
sub-samples of galaxies that occupy lower and higher mass halos. We use 
galaxy absolute magnitude to divide the overall sample into these two
subsamples, making the assumption that more luminous galaxies are associated with 
higher mass, and consequently higher bias halos \citet{Uitert2015}. 
We use five different approaches to divide the dataset into two sub-samples,
using magnitude in the five different SDSS bands. 

Our procedure is as follows: We first bin the sample into redshift bins with $ \Delta z = 0.01$. In each bin, we compute the median magnitude for one
of the five photometric bands ($u, g, r, i, z$). Galaxies brighter than
 the median magnitude in the bin are placed into the bright subsample
and fainter galaxies in the faint subsample. We repeat the process for
the other four photometric bands so that we have five different partitions of
the whole dataset into bright and faint subsamples. This process guarantees  that the two sub-samples have same 
redshift distribution. Our measurements are not 
sensitive to choice of $\Delta z$, we provide more details of this approach and other tests in section \ref{sec:Msys} and Figure \ref{fig:gz-sys}. 

The photometric band magnitudes used in our analysis are adjusted for evolution and the k-correction to $z=0.55$. We did not account for error on the magnitudes which will lead to a scatter in the galaxy around the dividing magnitude. The faint sub-sample contains 382,711 galaxies and the bright sub-sample contains 382,722 galaxies, when $r$ band
magnitudes are used to define the cut. Figure \ref{fig:subsample} shows the distribution of galaxies in right ascension and distance from the observer in these two CMASS subsamples.  We obtain similar samples using the other  photometric bands.

\subsection{Estimating the cross-correlation function}
\label{sec:2DCF}
We adopt as a fiducial cosmology a flat $\Lambda$CDM-GR cosmological model with $\Omega_m=0.274$, $H_0=0.7$ ,$\Omega_bh^2 =0.0224$, $n_s=0.95$ and $\sigma_8=0.8$  \citep{And14} in order to convert observed celestial coordinates ($\alpha,\delta$) and redshift to the position of the galaxy in 
three-dimensional space. These galaxy positions are used to estimate the two 
point statistic (cross-correlation function) of the galaxies in the
two subsamples using the minimum variance, Landy-Szalay estimator \citep{LandySzalay93}: 
\begin{align}
&\hat{\xi}(r,\theta) = \\
& \frac{D1D2( r,\theta) -D1R2(r,\theta) -R1D2(r,\theta)+R1R2(r,\theta)}{R1R2(r,\theta)} \nonumber
\end{align}
Here  $D1D2, D1R2, R1D2$ and $R1R2$ represent, respectively, 
the weighted counts of galaxy pairs from the two populations, pairs of galaxies in the first population with randoms for the second population, randoms for first population with galaxies from the second population and between randoms for two populations. We use the weighting scheme  $w_{sys}=w_{star} w_{see} (w_{cp}+w_{zf}-1)$  described in \citet{And14} to account for systematic weights of the individual galaxies. The weight factor $w_{zf}$ accounts for redshift failure of the nearest neighbor of a galaxy. The weight factor $w_{cp}$ is intended to account for a scenario where the redshift of a neighbor was not obtained because it was in a close pair due to fiber collision. The weights $w_{star}$ and $w_{see}$ account for stellar density and the seeing effect in the galaxy sample.

The cross-correlation function depends on $r$,  the distance between a pair of galaxies, and $\mu=cos(\theta)$, where $\theta$ is the angle between
the pair separation vector and the line of sight. We define the line-of-sight direction for each pair to be the position vector that joins the observer to the mean position of that pair of galaxies. We carry out cross-correlation function measurements covering  1 h$^{-1}$Mpc$<r<$ 60 h$^{-1}$Mpc with 15 logarithmic bins and $0<\theta<\pi$ with 150 linear bins. Provided that the binning is not much finer or coarser than these values our measurements are insensitive to binning choices.

\subsection{Estimating Multipoles and Galaxy bias}
\label{sec:mbias}
We measure the 2D cross-correlation function $\xi(r,\theta)$ from the CMASS data as described in section \ref{sec:2DCF}.  We compress the cross-correlation
 by projecting it onto a basis of Legendre polynomials $L_\ell(\mu)$ of order $\ell$ as follows:

\begin{align}
\hat{\xi}_\ell (r) & = \frac{2l+1}{2} \int_{-1}^1 d\mu \hat{\xi}(r,\mu)L_\ell(\mu) \\
& \approx \frac{2l+1}{2} \sum_{k} \Delta \mu_k \hat{\xi}(r,\mu) L_\ell(\mu_k),
\end{align}
where $\mu=\cos(\theta)$ . The $\ell=\{0,1,2\}$ moments of the Legendre polynomials are given by $L_\ell(\mu)=\{1,\mu,1/2(3\mu^2-1)\}$, the monopole, dipole and quadrupole, respectively. We estimate the linear bias $b$ of a sample 
of galaxies by fitting the model $\xi_0^{theo}=b^2 \xi_0^{m}$ to the observed monopole from data. Convolution Lagrangian Perturbation Theory  \citep[CLPT;][]{Carlson12} is used to estimate the monopole of the matter correlation function ($\xi_0^{m}$) assuming fiducial cosmology for $z=0.57$. We also estimate the dipole moment $\xi_{\ell=1}$ and use it as a means to detect asymmetry in the cross-correlation function \cite{Gaztanaga2015}.

\subsection{Estimating the Covariance Matrix}
\label{sec:covmat}

Estimation of the covariance matrix of a summary statistic (such as the cross-correlation function) is one of the most important steps in a cosmological analysis. The covariance matrix of an observed statistic is usually computed either by sub-sampling the data or by using mock catalogues. Both  methods have their limitations and regime of validity. Generally speaking, the sub-sampling methods over-estimate errors on small scale \citep{Norberg2009} and are difficult to use on large scales due to the limited volume of the observed data. Conversely creating realistic mock catalogues in large numbers and covering a large volume with high spatial resolution requires huge computing resources. Therefore, mocks often use approximate simulations with lower  spatial and temporal resolution \citep[e.g.][]{2014MNRAS.437.2594W}. This approach makes small scale clustering in the mocks inaccurate and hence covariance estimated from mocks can only be used above a certain minimum scale decided by details of the method. 

Because our signal of interest is on small scales ($r\sim 3-20 $h$^{-1}$Mpc)
we adopt the subsampling approach. We use  the ``delete one jackknife'' algorithm \citep{Shao1986} to estimate the covariance matrix. We first split the data into 210 approximately equal area regions (154 in the North Galactic cap and 56 in the South Galactic cap) as shown in Figure \ref{fig:jacknife}. A realization of data is defined by omitting one region at a time, which yields 210 realizations. We measure the summary statistics, correlation function and shell estimator for each realization. We then estimate the covariance matrix of these summary statistics ($ss$) using   

\begin{equation}
C_{i,j}=  \frac{N-1}{N} \sum_{jk=1}^{N} (ss_i^{jk}-\bar{ss}_i)(ss_j^{jk} -\bar{ss}_j).
\label{eq:jackcov}
\end{equation}
Here $C_{i,j}$ represents the covariance between bin $i$ and $j$, $\bar{ss}$ is the mean of the jackknife realizations and the sum is over all the 210 jackknife realizations. Our smallest jackknife region  has an angular diameter of $\sim 4^\circ$, which translates to $\sim 100$ h$^{-1}$Mpc. This distance is much larger than largest scale we are using in our analysis. 

\subsection{Shell estimator: Estimating Asymmetry}
\label{sec:shell}

As our primary measure of the redshift asymmetry in clustering
caused by relativistic effects, we use a shell-averaged estimator
applied to the  cross-correlation function. \citet{Croft2013} proposed this 
estimator to quantify the effects of gravitational redshift predicted from in $N$-body simulations. The shell estimator is defined as follows:
\begin{equation}
z_g^{shell}(s) =\frac{ \int_{\theta=0}^{\theta=\pi} H s_{\parallel} [1+\xi (s,\theta)]   d\theta}{ \int_{\theta=0}^{\theta=\pi} [1+\xi (s,\theta)]  d\theta} 
\label{eq:zg-rtheta}
\end{equation}
where $s_{\parallel}$ is the angle of pair separation from line of sight. We can see that $z_{g}^{\rm shell}$ measures the mean $s_\parallel$ weighted by the cross correlation function and is converted to km/s units through multiplication factor of H=100 (km/s) / (h$^{-1}$Mpc). Other quantifications of
the relativistic asymmetry in clustering have been proposed, such as the
imaginary part of the power spectrum \cite{McDonald2009}, 
the dipole \cite{Bonvin2014b}, and the anti-symmetric part of 
the cross-correlation function \cite{Irsic2016}. Here we focus on the shell 
estimator, but also measure the dipole and compare conclusions derived from both.

\begin{figure}
\begin{center}
\includegraphics[width=0.5\textwidth]{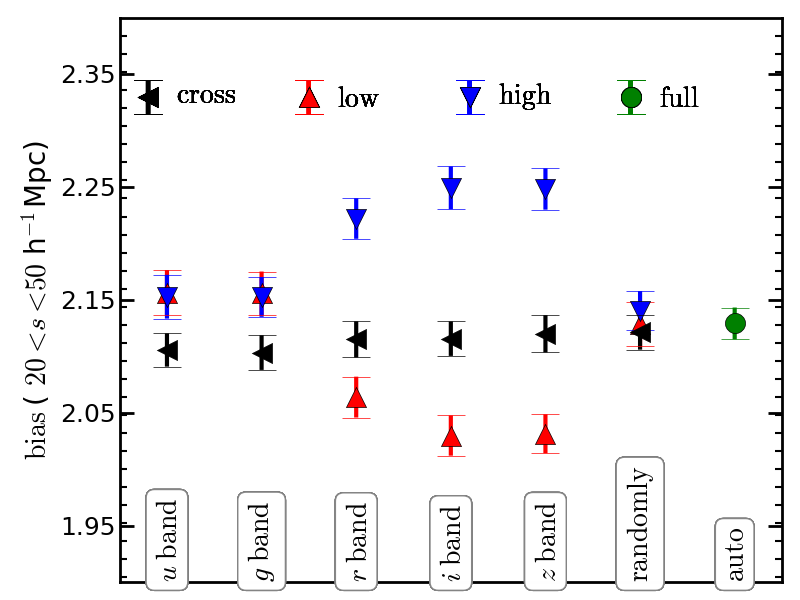}
\caption{The bias measured for each of the sub-samples used in our analysis using scales between 20 and 50 h$^{-1}$Mpc. The red, blue, black and green points represent bias of low-mass auto-correlation, high-mass auto-correlation, low-high cross-correlation and full sample auto-correlation functions, respectively. The subsamples are defined by each of the five photometric magnitudes ($u,g,r,i,z$) and also a random split. The $r,i$ and $z$ samples show significantly different biases for low and high mass subsamples.}
\label{fig:bias}
\end{center}
\end{figure}

\section{Measurements, Null Tests and Systematics}
\label{sec:result}

\begin{figure*}
\begin{center}
\includegraphics[width=0.95\textwidth]
{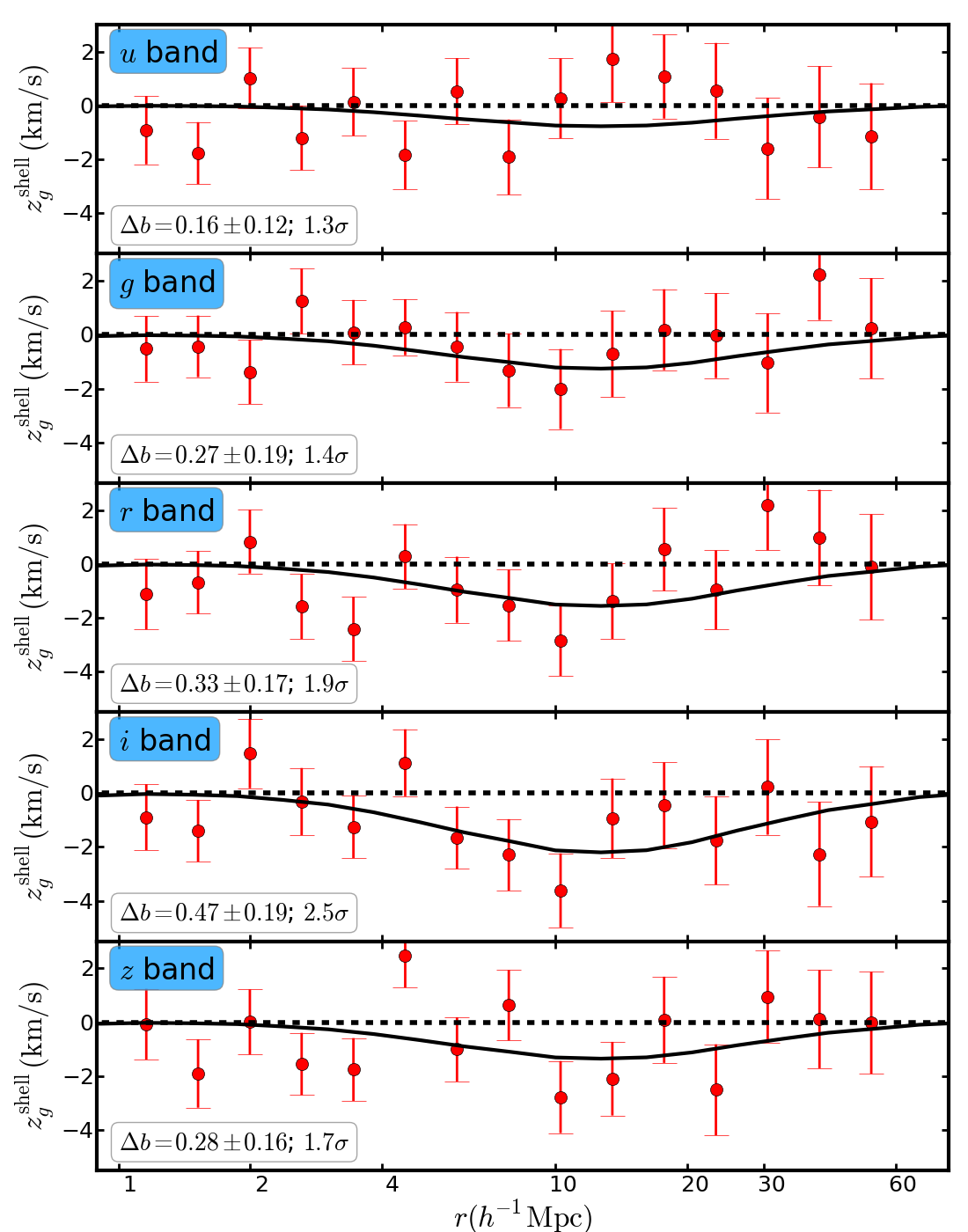}
\caption{The measurement of shell estimator from SDSS CMASS sample. The five different panels show the shell estimator measured using cross-correlation of sub-samples created by splitting the sample in two equal halves for each of $u,g,r,i,z$ photometric bands.  We detect the amplitude of relativistic asymmetry by measuring bias difference at  the level of $1.9\sigma$ ,$2.5\sigma$ and $1.7\sigma$ away from zero in the $r,i$ and $z$ bands, respectively. This result is consistent with our expectation from bias measurements of the five sub-samples given in Figure \ref{fig:bias}. The bias difference for $u$ and $g$ bands are at the level of $1.3\sigma$ and $1.4\sigma$, consistent with the expectation from biases.}
\label{fig:gz-measurement}
\end{center}
\end{figure*}

We use the methodologies described in previous section to perform our clustering measurements. In this section we present the measurement of bias, the measurements of shell estimator and fitting the model described in section \ref{sec:theory}. We also perform two null tests and look at the sensitivity of our observed signal to the possible systematic.

\subsection{Measurements of Bias}
\label{sec:Mbias}
The linear biases measured for our various samples are displayed in Figure \ref{fig:bias}. The galaxy bias is measured using the monopole $\xi_0(s)$ as described in section \ref{sec:mbias}. We use scales between 20h$^{-1}$Mpc and 50h$^{-1}$Mpc to measure bias. The bias of auto-correlation and cross-correlation of different sub-samples created by splitting the sample using all five photometric bands ($u,g,r,i,z$) and randomly are presented in the figure. The blue points show the biases of higher mass samples, red points for lower mass samples and the black points for the cross-correlation between the lower and higher mass samples. The bias of full sample is indicated by the green point. The relativistic effects, dominantly the gravitational redshift, breaks the line of sight symmetry of cross-correlation and is proportional to the difference in biases of the two sub-samples. Therefore we expect to see relatively smaller signal for $u$ and $g$ bands and relatively larger signal for $r,i$ and $z$ bands. We also expect no line of sight asymmetry in the cross-correlation using random split and the auto-correlation of the full sample. These two cases are used as our null tests.

\subsection{Measurements of the shell estimator}
\label{sec:Mshell}
Figure \ref{fig:gz-measurement} presents the measurements of the shell
estimator together with the
 best fit model. The red points show our measurement of the asymmetry
which we interpret as being due to
relativistic effects including gravitational redshift ($z_g^{\rm
  shell}(r)$). The solid black lines indicate the expected signal
based on our best fit model. We have used predictions from $N$-body
simulation to fit the measurements for each band using the model described
in equation \ref{eq:gz-model}. Our likelihood function is defined as
follows

\begin{align}
& \mathcal{L}(\Delta b)= e^{-\chi^2(\Delta b)/2} / \int
  e^{-\chi^2(\Delta b)/2} d\Delta b\\ & \chi^2(\Delta b) =
  (z_g^{data}-z_g^{model}) C^{-1} (z_g^{data}-z_g^{model})^T
\end{align}
where $C^{-1}$ represents the inverse of the covariance matrix obtained
from jacknife sampling as shown in equation \ref{eq:jackcov}. The
likelihood is used to estimate the mean and error on 
$\Delta b$ measured from the model fit. This value of $\Delta b$
is therefore the one inferred from the clustering asymmetry, assuming 
relativistic effects. It can be compared to the value of $\Delta b$
directly measured from the ratio of the galaxy correlation functions,
to see if they are consistent.

 The requirement that the amplitude of the line-of-sight asymmetry
approaches zero ($A_{\rm rel} \to 0$) in the absence of any
relativistic effects is our null hypothesis. The detection
significance quantifies the difference of bias difference from zero in
units of standard error ($N_\sigma=\Delta b^{\mu}/\Delta b^\sigma$). We
fit only for the amplitude of asymmetry as the function of $\Delta b$.
Table \ref{tb:gz-fit} provides the results of our fits for each of the
photometric band. The best fit model is shown in Figure
\ref{fig:gz-measurement} with the black solid line. The measured bias
difference ($\Delta b$), its error and detection significance, are
also given in Figure \ref{fig:gz-measurement}.

\begin{table}
\caption{The best fit bias difference ($\Delta b$) for each of the
  shell estimator measurements. The table also lists the detection
  significance ($N\sigma$) for each measurement. The values in square
  brackets in the $\Delta b$ column are the measurement of linear bias
  difference using the monopole of the auto-correlation for the low
  and high bias samples (see section \ref{sec:mbias} for more
  details). The first row ${\rm`` All"}$ corresponds to fitting shell estimator from all five band simultaneously.}
\begin{center}
  \begin{tabular}{llll}
    ${\rm split}$ & $\Delta b$  & $N\sigma$ \\ \hline \hline
    ${\rm All}$ & $0.44\pm0.16$  & $2.7$\\
\hline
    $u$ & $0.16\pm0.12$ [0.0] & $1.3$\\
\hline
    $g$ & $0.27\pm0.19$ [0.0] &  $1.4$\\
     \hline
    $r$ & $0.33\pm0.17$ [0.16] &  $1.9$\\
 \hline
    $i$ & $0.47\pm0.19$ [0.23] & $2.5$\\
 \hline
    $z$ & $0.28 \pm 0.16$ [0.21] & $1.7$\\
 \hline \hline
    ${\rm random}$ & $0.08\pm0.08$ [0.01] &  $1.0$\\
  \hline
    ${\rm auto}$  & $0.01\pm0.01$ & $1.0$\\
  \hline
    \hline
  \end{tabular}
\end{center}
\label{tb:gz-fit}
\end{table}

The $u$ and $g$ bands show an amplitude of gravitational redshift
consistent with zero, which is expected given that the bias difference
of the two sub-samples using these bands is small. The $r,i$ and $z$
bands have a deviation of $A_{\rm rel}$ from zero at the
$1.9\sigma,2.5\sigma$ and $1.7\sigma$ levels, respectively. 
If we compare the bias
difference measured from the shell estimator (largely 
from non-linear scales) and from
monopole of the correlation function we find consistency at the $1.5\sigma$
level or better. 

\begin{figure}
\begin{center}
\includegraphics[width=0.5\textwidth]{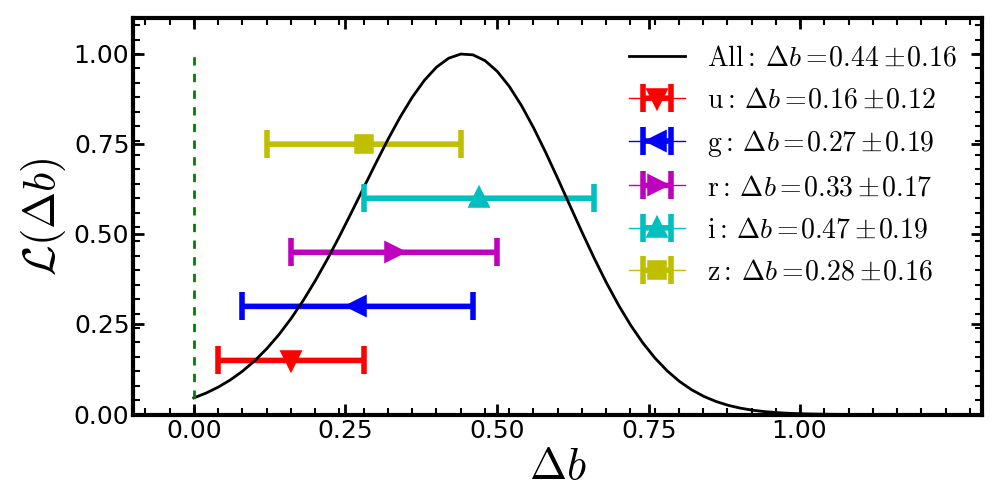}\\
\includegraphics[width=0.5\textwidth]{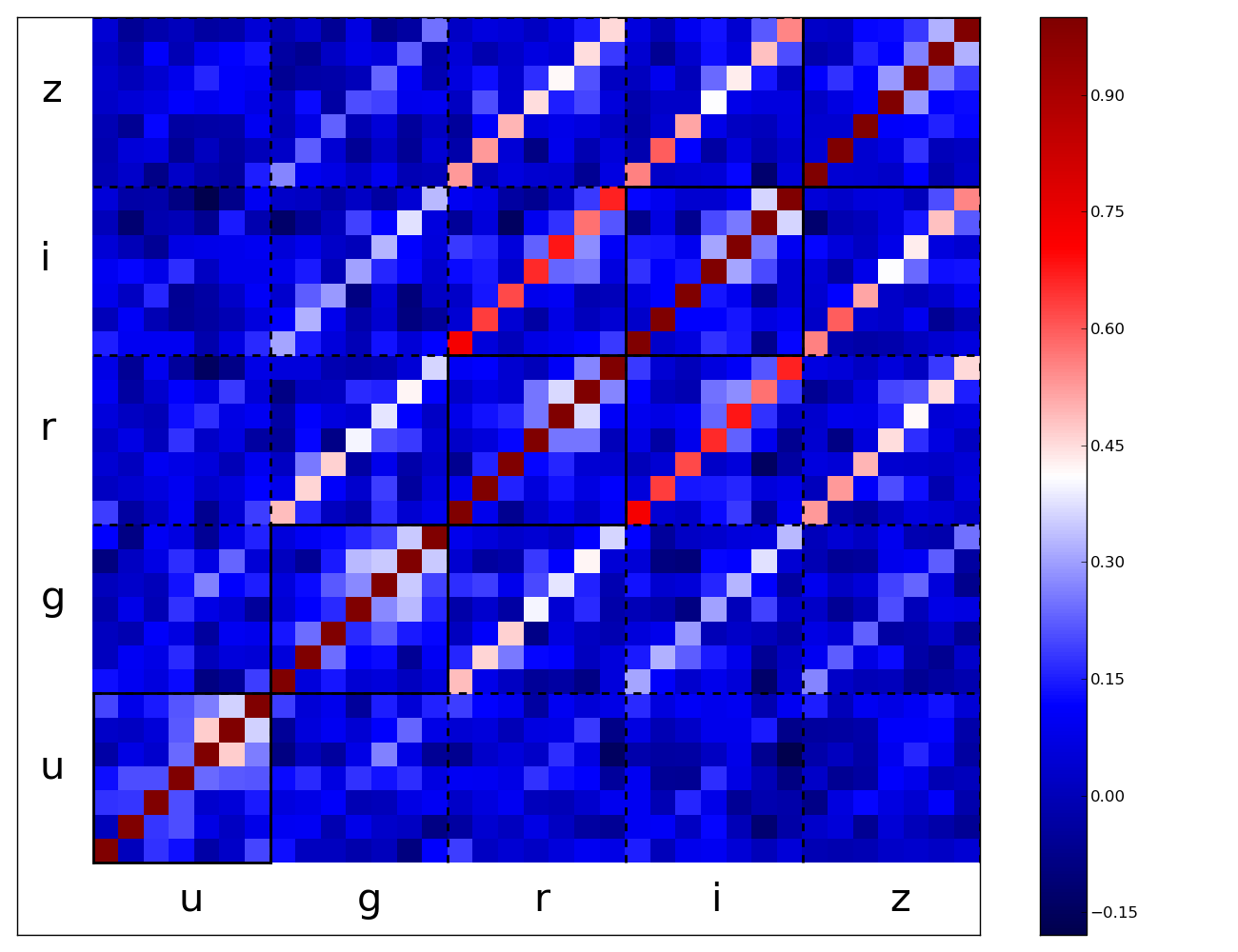}
\caption{The top plot shows the likelihood associated with measurement of
 $\Delta b$ from the clustering asymmetry (shell estimator),
   using data from the central 7 $r$ bins of all 5 bands
taken together. The
  different coloured markers denote the measurements from individual bands
  using all 15 bins. The bottom plot shows the normalised covariance matrix of
the  shell estimator for all five photometric bands. We use (and show) only the
  central 7 $r$ bins in order to avoid dominating the covariance matrix
  with noise. This matrix was  used to perform the simultaneous fit to
 the shell
  estimator for all five bands (curve in top plot). }
\label{fig:like-crossband}
\end{center}
\end{figure}

We have also performed a simultaneous fit to all five bands accounting
for the cross-correlation between the bands. We fit for one $\Delta b$
corresponding to $i$ band and scale the models for the other bands by
the ratio of the best fit $\Delta b$ of the band to that for
the $i$ band. In
order to be able to compute the covariance matrix among different
bands we use only the central 7 $r$ bins of the measured
$z_{\rm shell}$ values, covering scales
between 3 to 25 h$^{-1}$Mpc. This
combined measurement yields  a 2.7$\sigma$ detection
with $\Delta b=0.44 \pm 0.16$. Figure \ref{fig:like-crossband} shows
the likelihood and correlation matrix. The black line in the top plot
shows the likelihood when combining all 5 bands and the individual
points shows the measurements for the individual band fits. The bottom plot
shows the normalised covariance matrix within the band and across the band.
We can see that as expected there is strong covariance, particularly
between bins at the same scale in different bands (up to $\sim 0.7$
covariance). As a result there is only marginal improvement in the significance
of the measurement from combining the different bands.

\subsection{Null Tests}
\label{sec:Ntest}
\begin{figure}
\begin{center}
\includegraphics[width=0.5\textwidth]{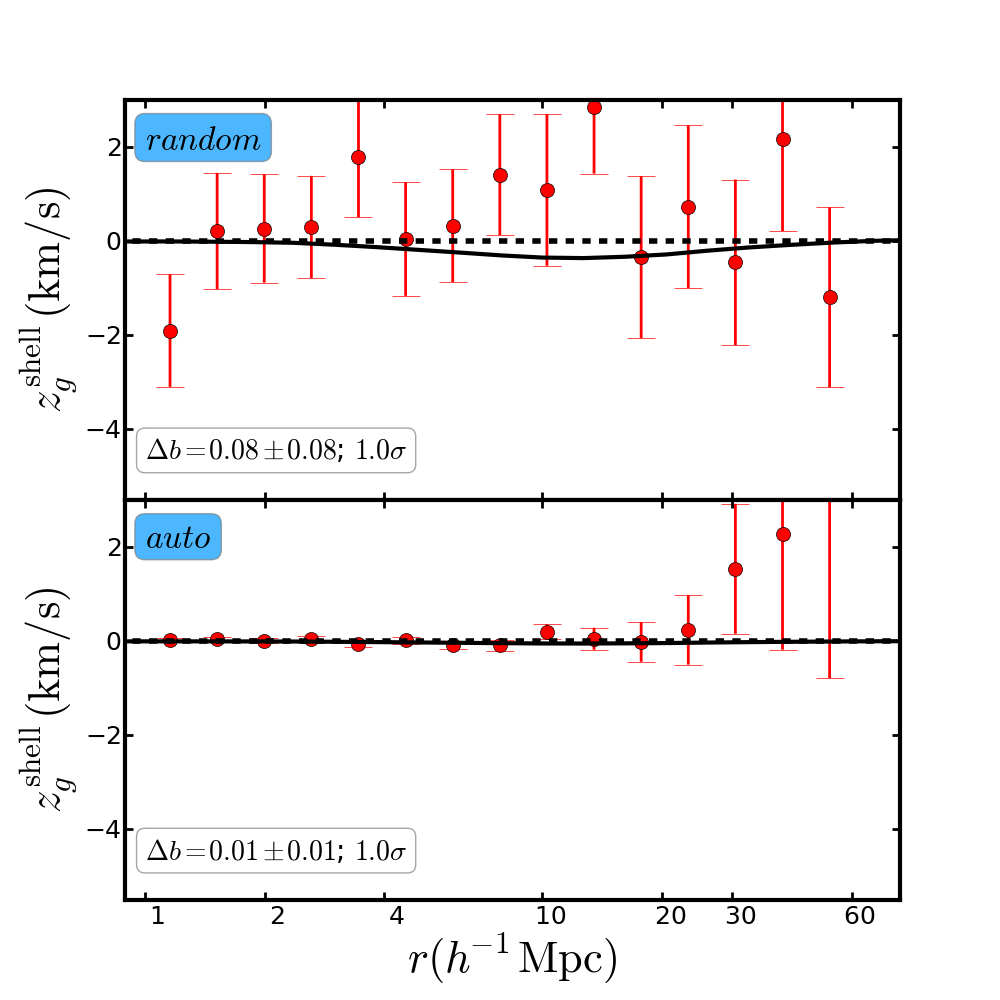}
\caption{The results of our null test of zero signal to check our pipeline and various possible systematic effects.  The top panel shows the shell estimator computed from cross-correlation when we split the sample randomly. The best fit signal amplitude is completely consistent with zero. The bottom panel shows the shell estimator computed from the auto-correlation of the full sample. Because this is an auto-correlation we do not expect to see any signal in the shell estimator. The plots demonstrate that we pass both the null tests because the best fit signals are consistent with zero.
}
\label{fig:gz-nulltest}
\end{center}
\end{figure}

We perform two null tests to check for systematics. First, we divide
the sample randomly in two equal halves and examine the shell
estimator from the cross-correlation of the two random sub-samples. We
do not expect to observe any signal showing line of sight asymmetry
from such a measurement, because the two sub-samples are statistically
same. The top panel in Figure \ref{fig:gz-nulltest} shows the shell
estimator measurement from the random split. We obtained $\Delta b=
0.08 \pm 0.08$ which is consistent with zero signal at the $1\sigma$
level. We have also investigated the shell estimator from the
auto-correlation of the full sample, which serves as the second null
test. The bottom panel in Figure \ref{fig:gz-nulltest} shows the
measurement of the shell estimator from the auto-correlation of the full
sample. This analysis yields $\Delta b= 0.01 \pm 0.01$, which is
consistent with zero at the $1\sigma$ level. Any problems with the survey
geometry, mask, large angle effect or redshift distribution should
produce a non-zero signal in at least one of these measurements. Note
that null test using the auto-correlation has a much smaller error bar and
still produces a null result which is a strong test for many of the
possible systematic effects. Both of our null tests are in excellent
agreement and suggest that our analysis is not affected with
significant systematic effects.

\begin{figure}
\begin{center}
\includegraphics[width=0.45\textwidth]{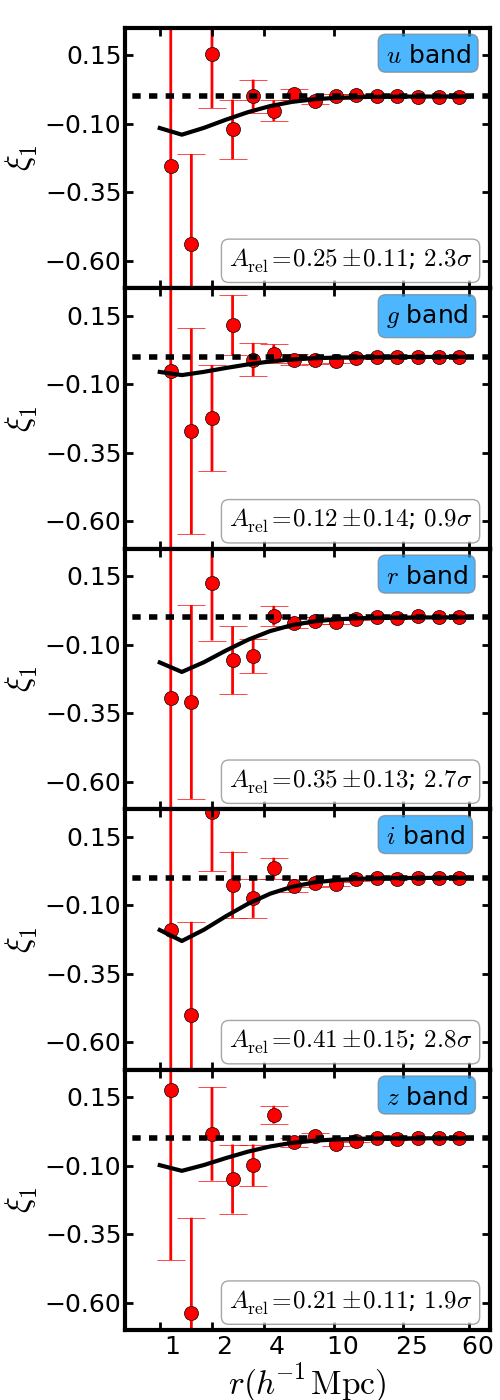}
\caption{The dipole moment measured using cross-correlation of sub-samples created by splitting the sample in two equal halves for each of $u,g,r,i,z$ photometric bands.  We detect the amplitude of relativistic asymmetry at  the level of $2.3\sigma$ ,$0.9\sigma$, $2.7\sigma$, $2.8\sigma$ and $1.9\sigma$  in the $u,g,r,i$ and $z$ bands, respectively. }
\label{fig:xi1-data}
\end{center}
\end{figure}

\subsection{Measurement of Dipole moment}
\label{sec:Mdipole}
We have measured the dipole moment of the cross-correlation for each of the
photometric band as described in section \ref{sec:mbias}; Figure
\ref{fig:xi1-data} presents these results. Each panel shows the dipole
moment of the cross-correlation for each of the photometric bands. The
red points indicate our measurement with a jackknife error bar. The
black line represents the best fit halo model prediction. The
reader is referred  to \citet{Croft2013} for details of halo model used for the
dipole moment, and for evidence that the model reproduces simulation
results. Briefly, the model estimates the mass function as the
integral of the difference between galaxy matter correlation function for
the two sub-samples and uses the mass function to estimate the
gravitational redshift in the context of general relativity (see
equations 2 and 3 of \citet{Croft2013}). 
The results shown here for the dipole are included to show that asymmetry
can be measured using a different statistic than the shell estimator.
We present the shell estimator as our primary result because in our 
companion paper   (\citep[][in press.]{Zhu2016Nbody}) we found making simulation
predictions for the dipole including all relativistic effects to be
technically unfeasible.

 We fit the halo model
prediction for a constant amplitude which is a multiplicative factor
to our fiducial halo model prediction. The dipole moment shows a
non-zero signal at small scales similar to the shell estimator.  
The figure also shows the best fit
value of the halo model amplitude. We detect the amplitude of
relativistic asymmetry at the level of $2.3\sigma$ ,$0.9\sigma$,
$2.7\sigma$, $2.8\sigma$ and $1.9\sigma$ in the $u,g,r,i$ and $z$
bands, respectively, but do not detect any dipole signal at large
scale. The results are entirely consistent with zero for scales above
25h$^{-1}$Mpc.

\subsection{Sensitivity to Systematics}
\label{sec:Msys}
\begin{figure*}
\begin{center}
\includegraphics[width=1.0\textwidth]{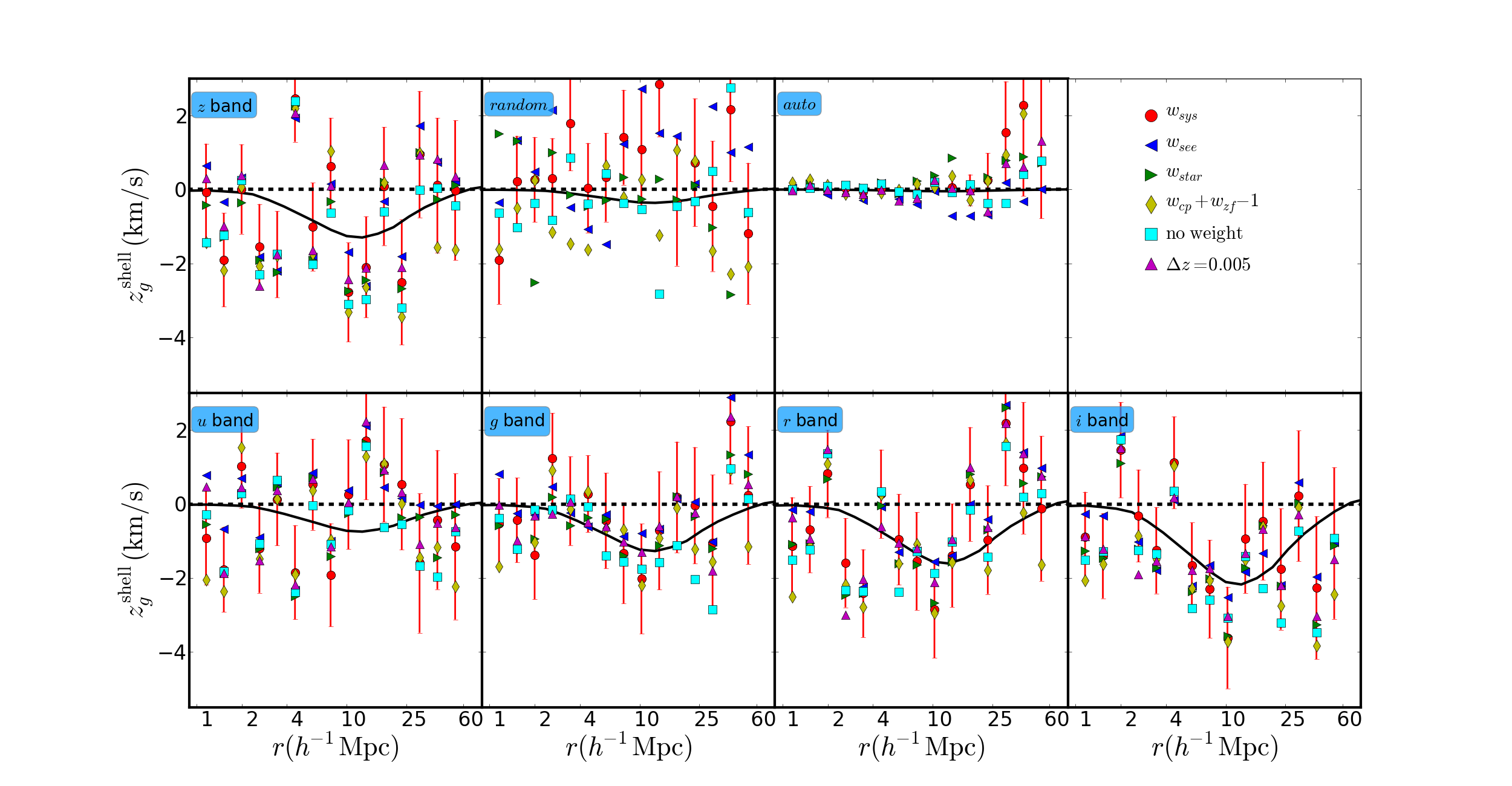}
\caption{This plot shows the effect of different systematic weights on the measurement of shell estimator for each of the five photometric bands and the two null tests. The highlight here is that our measurement is not very sensitive to the choice of systematic weights or the width of redshift bin used while creating our sub-samples. The different coloured points are when we include different systematic weights.}
\label{fig:gz-sys}
\end{center}
\end{figure*}

We have used the systematic weights suggested in the catalog in an
attempt to remove any observational correlation that shouldn't exist
in our samples. We apply five combinations of systematic weights, as
listed in the legend of Figure \ref{fig:gz-sys}, in the measurement of
cross-correlation for each of the photometric bands and null tests. We
compute the shell estimator for each of these weights for each
sub-samples, produces 35 different measurements, which are all
displayed in Figure \ref{fig:gz-sys}. The seven different panels show
the shell estimator measurement for each photometric bands and also
the ones used for our null-tests. The coloured points represent
different combinations of systematic weights used in the correlation
function measurement. The red points in each panel are our fiducial
measurement. We found that the effect of any of these systematic
weights are small, and obtain consistent measurement within $1\sigma$
independent of what systematic weights are used. This exercise
demonstrate that our measurements are robust against observational
systematic. We have also tested the effect of redshift binning while
creating the sub-samples. Our fiducial bin width is $\Delta z =0.01$,
as shown using red points. Decreasing the bin width to $\Delta z
=0.005$, as shown using magenta points, does not change the
measurement significantly. The fact that our measurements is not
sensitive to the combination of weights used or choice of redshift
binning suggests that it may be stable against lack of detailed
understanding of some of these systematic weights.

\section{Discussion}
\label{sec:discussion}
We have discussed various relativistic effects which could produce
line-of-sight asymmetries in the cross-correlation of two galaxy
populations with different mean halo mass and used the BOSS CMASS sample to
measure these asymmetries. We used each of the five SDSS photometric
bands in turn to obtain five sets of 
two sub-samples. The galaxy biases of these
sub-samples were measured by measuring the monopole of auto- and
cross-correlation as described in section \ref{sec:mbias} and shown in
Figure \ref{fig:bias}. The shell estimator described in section
\ref{sec:shell} was used to quantify the line-of-sight anisotropy due
to the relativistic effects in velocity units. We have used a model
developed using $N$-body simulations and motivated by perturbation
theory to fit the measured shell-estimator. The theoretical model for
shell estimator is described in section \ref{sec:theory} and derived
from the analysis of a companion paper \citet[][in press.]{Zhu2016Nbody}. The
theoretical model was fit to measurements from data in order to
quantify the significance of observed signal. The covariance matrix
which estimates the uncertainties was obtained using jackknife
sampling as described in section \ref{sec:covmat}. Figure
\ref{fig:gz-measurement} shows our measurements with best fit models
and detection significance. We detect the amplitude of relativistic
asymmetry at the level of $1.9\sigma$, $2.5\sigma$ and $1.7\sigma$
away from zero in the $r,i$ and $z$ bands respectively. This result is
consistent with our expectation from the relative bias measurements of
the five sub-samples as shown in Figure \ref{fig:bias}. The amplitude
of relativistic asymmetry for the $u$ and $g$ bands are at the level
of $1.3\sigma$ and $1.4\sigma$, consistent with the expectation from
biases. We detect the signal at 2.7$\sigma$ after combining the
measurements from all 5 bands including the cross-covariance (using only
the central 7 bins covering scales between 3 to 25 h$^{-1}$Mpc.)

\citet{Zhu2016Nbody} found that the dominant contribution in the shell
estimator is due to the gravitational redshift effects. Therefore our
measured signal can  be seen as a first detection of gravitational redshifts
distorting the large scale structure of the Universe. Two null tests
were devised to check the possibility that the measured line-of-sight
asymmetry is due to the survey geometry, mask effects, large angle
effects, redshift distribution, etc. The first null test measures the
shell estimator from the cross-correlation of two randomly selected
galaxy populations. This procedure makes the two populations
statistically identical and hence we do not expect any
line-of-sight-anisotropy in the cross-correlation. The second null
test uses the shell estimator measured from the auto-correlation of
the full sample. This test has much smaller error, allowing to check
for possible geometrical effects to much higher precision than the
statistical uncertainty in our measurement. Figure \ref{fig:gz-sys}
demonstrates that our null tests are consistent with zero, implying
that any of those possible systematic effects lie significantly below
our statistical uncertainty. We have also performed a much more
detailed analysis of possible systematic effects due to sample
selection in the companion paper \citet[][in press.]{Alam2016TS}.

Another avenue to quantify the line-of-sight asymmetry is the dipole
moment of the cross-correlation function (described in section
\ref{sec:mbias} and shown in Figure \ref{fig:xi1-data}).  We found
similar levels of asymmetry on small scales to those found with the
shell estimator. We fit our dipole moments using the prediction from
halo model described in \citet{Croft2013}, and detect the amplitude of
relativistic asymmetry at the significance level of $2.3\sigma$
,$0.9\sigma$, $2.7\sigma$, $2.8\sigma$ and $1.9\sigma$ in the
$u,g,r,i$ and $z$ bands respectively. 
Differences between the two measurements are likely
caused by the fact that the shell
estimator and dipole moment weight the clustering 
modes differently.  Our dipole
moments are completely consistent with zero on large scales. This
result is also consistent with the analysis presented by
\citet{Gaztanaga2015}, who  did not detect any signal using
dipole moment and shell estimator on large scales. We used a halo
model to measure the dipole moment signal; this model is not as
rigorously tested as our model for the shell estimator based on
simulations. A better understanding of dipole moment models based on
analytical methods or simulation will be a necessary step for the
future work on the measurement of relativistic asymmetry.

This first detection of gravitational redshifts using large scale
structure can be regarded another success for General Relativity, however, 
 we have found in companion work
 \citep[including ][in press.]{Zhu2016Nbody} that there are significant uncertainties in the
theoretical predictions for relativistic distortions, particularly
related to the role of structure on galactic scales. The observational
measurements, from large-scale galaxy clustering in this paper, and
from galaxy clusters \citep[e.g. ][]{Wojtak2011, Sadeh2015} are also
currently at only the 2-3 $\sigma$ of significance. Future surveys
with greater volume and larger numbers of galaxies will be able to
detect such signals with much higher significance. Such
surveys  will not only look at much larger volumes
but also probe much deeper, covering a wider variety of galaxies and
greater differences in the relative biases of samples. These surveys,
including ongoing programs such as eBOSS and future projects such as DESI
and Euclid, are already developing large
simulations and suites of analysis tools which will be available for
relativistic clustering studies. These efforts have
 the potential to bring these
distortions of large-scale structure into the realm of precision
cosmology and allow unprecedented tests of General Relativity.

\section*{Acknowledgments}
We would like to thank Nick Kaiser for many insightful discussions during the course of this project. This work was supported by NSF grant AST1412966.  SA and SH are supported by NASA grants 12-EUCLID11-0004. SA is also supported by the European
Research Council through the COSFORM Research Grant(\#670193). We would like to thank Ayesha Fatima for going through the early draft and helping us making the text much more clear. 

SDSS-III is managed by the Astrophysical Research Consortium for the Participating Institutions of the SDSS-III Collaboration including the University of Arizona, the Brazilian Participation Group, Brookhaven National Laboratory, Carnegie Mellon University, University of Florida, the French Participation Group, the German Participation Group, Harvard University, the Instituto de As trofisica de Canarias, the Michigan State/Notre Dame/JINA Participation Group, Johns Hopkins University, Lawrence Berkeley National Laboratory, Max Planck Institute for Astrophysics, Max Planck Institute for Extraterrestrial Physics, New Mexico State University, New York University, Ohio State University, Pennsylvania State University, University of Portsmouth, Princeton University, the Spanish Participation Group, University of Tokyo, University of Utah, Vanderbilt University, University of Virginia, University of Washington, and Yale University.

\bibliography{../Master_Shadab.bib}
\bibliographystyle{mnras}

\label{lastpage}

\end{document}